\documentstyle[12pt,fleqn]{article}
\begin{document} 

\title{Two Open Universes Connected by a Wormhole:\\Exact Solutions}
\author{Li-Xin Li\\
Department of
Astrophysical Sciences, Princeton University\\ Princeton, NJ 08544, USA}
\date{September 28, 2000}
\maketitle

\begin{abstract}
In this paper I present a spacetime of two open universes connected by a 
Lorentzian wormhole. The spacetime has the following features: (1) It
can exactly solve the Einstein equations; (2) The weak energy condition is
satisfied everywhere; (3) It has a topology of $R^2\times T_g$ ($g\ge 2$);
(4) It has no event horizons.

\end{abstract}

\newpage
Wormholes (or quantum foams) could play important roles on both microscopic and
macroscopic scales in the realm of classical and quantum gravity 
\cite{whe55,mis57,haw88,gid88}. Field lines
trapped in wormholes have led to the concepts of ``mass without mass'' and ``charge
without charge'' \cite{whe55,mis57}. In the context of Euclidean quantum gravity,
Euclidean wormholes arise from topology change on the Planckian scales, which have 
been proposed to cause the loss of quantum coherence or to fix coupling constants
\cite{haw88,gid88}. (However, see the recent
discussions of Hawking \cite{haw96}.) More interestingly, Morris, Thorne, and
Yurtsever \cite{mor88} (see also Novikov \cite{fro90}) have shown that a Lorentzian
wormhole can be transformed into a time machine (i.e., a spacetime with closed
timelike curves). Morris and Thorne \cite{mor88a}
have also considered the possibility of using Lorentzian wormholes as tools for
interstellar travel. Applications of wormholes in cosmology have also been
investigated \cite{hoc93,liu95}.

In this paper I focus on Lorentzian wormholes, which are essentially the
time-development of three-dimensional wormholes \cite{mtw73,vis95}.
Such kind of spacetimes exhibit wormhole structures on their spacelike foliations,
the wormholes either connect one universe to another universe
or connect one region to another distant region in the same universe. As far as I
am aware, almost all of the Lorentzian wormhole solutions discovered in
literatures have two-dimensional spacelike cross-sections with a topology of $S^2$.
With very generic arguments, Morris and Thorne \cite{mor88a}
have shown that for static, spherically symmetric, and traversable (i.e., having no
event horizons) wormholes the weak energy condition must be violated near the
wormhole throats. Dynamic (evolving) wormholes have also been
considered \cite{kar94,kar96,rom93,rom94,wan94,anc98}. It has been demonstrated that
dynamic wormholes without violation of the weak energy condition could exist within
a finite period of time \cite{kar94,kar96}. In this paper, I present a
wormhole spacetime whose two-dimensional spacelike cross-sections have a topology of
$T_g$ ($g\ge 2$) instead of $S^2$ that is usually considered in literatures, where
$T_g$ is a $g$-torus. I show that, the wormhole spacetime represents two open
universes connected by a Lorentzian wormhole and has the following features:
(1) It can exactly solve the Einstein equations; (2) The weak energy condition is
satisfied everywhere; (3) It has a topology of $R^2\times T_g$ ($g\ge 2$);
(4) It has no event horizons.

The wormhole spacetime is constructed from a usual open Friedmann-Robertson-Walker
(FRW) universe. An open FRW universe has negative spatial curvature and can be
foliated with spatial hyperbolic hypersurfaces ($H^3$). Each $H^3$ can be embedded
in a four-dimensional Minkowski spacetime through \cite{haw73}
\begin{eqnarray}
    T^2-X^2-Y^2-Z^2 = a^2,
    \label{h3}
\end{eqnarray}
where $(T,X,Y,Z)$ are the Cartesian coordinates in the Minkowski spacetime, $a$ is
a constant. (Throughout the paper I use the geometric units $G = c =1$.) The metric
of the Minkowski spacetime is
\begin{eqnarray}
    dS^2 = - dT^2 + dX^2 + dY^2 + dZ^2\,.
    \label{min}
\end{eqnarray}
Different $H^3$ can have a different radius, thus $a$ is a function of the cosmic
time $t$ which labels the spacelike foliation of the universe.
Let's cut an $H^3$ at $Z = {\rm constant}$, then we obtain a two-dimensional
hyperbolic surface $H^2$ with a radius $(a^2+Z^2)^{1/2}$. An $H^2$ is a
two-dimensional Lobachevskii space, which can be compactified through
$H^2/\Gamma$, where $\Gamma$ is an isometric group without fixed points on $H^2$.
If we require manifolds to be orientable, then the only compact surfaces
that can be obtained from $H^2$ are $g$-tori $T_g$ with $g \ge 2$ being the
genus (i.e., the number of holes) of the surfaces \cite{lac95,ben92,got80}.
A $T_g$ with $g\ge 1$
can be obtained as the quotient of a polygon of $4g$ sides with all vertices being
identified with each other and the sides being identified in pairs \cite{ben92}.
Now let's start with an open FRW universe and compactify its two-dimensional
cross-section $H^2$ to a $T_g$ ($g\ge 2$), then
we obtain a spacetime $(M,g_{ab})$ which has the same metric as an ordinary
open FRW universe but has a topology of $M=R^2\times T_g$ ($g\ge 2$).
[An ordinary open FRW universe, i.e. the covering space of $(M,g_{ab})$, has a
topology of $R\times H^3$.] I will show that the spacetime $(M,g_{ab})$ represents
two open universes connected by an evolving Lorentzian wormhole.

The spacetime $(M,g_{ab})$ can be covered locally with coordinates
$(t,\chi,\xi,\phi)$, where $t$ is the cosmic time and $(\chi,\xi,\phi)$
are defined with $(T,X,Y,Z)$ through
\begin{eqnarray}
    T &=& a \cosh\chi\cosh\xi\,,
    \label{cot}\\
    X &=& a \cosh\chi\sinh\xi\cos\phi\,,
    \label{cox}\\
    Y &=& a \cosh\chi\sinh\xi\sin\phi\,,
    \label{coy}\\
    Z &=& a \sinh\chi\,.
    \label{coz}
\end{eqnarray}
Then the FRW metric is given by
\begin{eqnarray}
    ds^2 = -dt^2 + a(t)^2\left[d\chi^2 + \cosh^2\chi \left(d\xi^2 +
           \sinh^2\xi ~d\phi^2\right)\right].
    \label{met}
\end{eqnarray}
Consider the two-dimensional cross-section ${\cal S} = T_g (g\ge 2)$ with
$t = {\rm constant}$ and $\chi = {\rm constant}$. The metric on ${\cal S}$ is
\begin{eqnarray}
    d\overline{s}^2 = a^2 \cosh^2\chi\,\left(d\xi^2 +
    \sinh^2\xi ~d\phi^2\right)\,.
    \label{met1}
\end{eqnarray}
According to the Gauss-Bonnet theorem, the area of ${\cal S}$ is \cite{ben92,ber81}
\begin{eqnarray}
    {\cal A} = 4\pi (g-1) a^2 \cosh^2\chi\,.
    \label{area}
\end{eqnarray}
Thus, the size of ${\cal S}$ is completely determined by its genus and the metric on it.
On the hypersurface $t = {\rm constant}$, ${\cal A}$ takes the minimum at $\chi = 0$:
${\cal A}_{\min} = 4\pi (g-1) a^2$ which is non-zero for $g\ge 2$. As $\chi\rightarrow
\pm\infty$,
${\cal A}\rightarrow\infty$ exponentially and we enter two open universes (they
are open since the hypersurface $t = {\rm constant}$ has uniform negative curvature
everywhere). Therefore, $(M,g_{ab})$ represents a spacetime of two open universes connected
by a wormhole. One open universe is on the side with $\chi>0$, the other is on the
side with $\chi<0$. The ``throat" of the wormhole is at $\chi = 0$. Since $a = a(t)$, the
wormhole evolves with time.

The coordinate $\chi$ measures the proper distance from the wormhole's throat
on the hypersurface $t = {\rm constant}$, which goes from $-\infty$ to $\infty$.
Define $r  = \cosh\chi$, then the metric in Eq.~(\ref{met}) can be written as
\begin{eqnarray}
    ds^2 = -dt^2 + a(t)^2\left[{dr^2\over r^2-1} + r^2 \left(d\xi^2 +
           \sinh^2\xi ~d\phi^2\right)\right].
    \label{met2}
\end{eqnarray}
However, $r>1$ covers only one side of the wormhole: either the region with
$\chi>0$, or the region with $\chi<0$.

Since $(M,g_{ab})$ is locally isometric to an open FRW universe, it
can exactly solve the Einstein equations with a suitable stress-energy tensor.
Let's take the stress-energy tensor to be that of a perfect fluid: $T_{ab} =
(\rho+p)u_a u_b + p g_{ab}$, where $u^a = (\partial/\partial t)^a$ is the
four-velocity of comoving observers, $\rho = \rho(t)$ and $p = p(t)$ are the energy
density and pressure measured by comoving observers.
Then for the metric in Eq.~(\ref{met}) the Einstein equations are reduced to the
familiar forms for an open FRW universe
\begin{eqnarray}
    \left(\dot{a}\over a\right)^2 &=& {8\pi\over 3} \rho + {1\over a^2} +
    {\Lambda\over 3}\,,
    \label{eq1}\\
    {\ddot{a}\over a} &=& -{4\pi\over 3}\left(\rho+3p\right) +
    {\Lambda\over 3}\,,
    \label{eq2}
\end{eqnarray}
where the dot denotes $d/dt$, $\Lambda$ is the cosmological constant. Supplied with
a state equation $p = p(\rho)$, Eq.~(\ref{eq1}) and Eq.~(\ref{eq2}) can be solved.
There are many exact solutions, here I list some of them \cite{pee93}:\\
{\it 1) Vacuum solution:} Let $\rho = p = \Lambda = 0$, then we have the simplest
solution
\begin{eqnarray}
   a = t,
   \label{vac}
\end{eqnarray}
which is locally a Milne universe.\\
{\it 2) Wormhole supported by dust:} In such a case $p = 0$,
$\Lambda = 0$, and $\rho a^3 = {\rm constant}\equiv 3C_1/8\pi$. The solutions are
\begin{eqnarray}
    a = {1\over 2}C_1\left(\cosh\eta - 1\right), \hspace{1cm}
    t = {1\over 2}C_1\left(\sinh\eta - \eta\right),
    \label{dus}
\end{eqnarray}
where $\eta>0$ is a parameter.\\
{\it 3) Wormhole supported by radiation:} In such a case $p = \rho/3$,
$\Lambda = 0$, and $\rho a^4 = {\rm constant}\equiv 3C_2/8\pi$. The solutions are
\begin{eqnarray}
    a = \left(t^2 + 2 C_2^{1/2} t\right)^{1/2}.
    \label{rad}
\end{eqnarray}
{\it 4) Wormhole supported by a cosmological constant:} Suppose $\Lambda>0$
and $\rho = p = 0$. Then we have the open de~Sitter solutions
\begin{eqnarray}
    a = \left(3\over\Lambda\right)^{1/2}\sinh\left[
    \left(\Lambda\over 3\right)^{1/2}t\right].
    \label{des}
\end{eqnarray}
These are inflating wormholes.

Since all these wormhole solutions are supported by ordinary matter (dust, radiation,
or simply vacuum) or a positive cosmological constant, the weak energy condition
(which requires $\rho\ge 0$ and $\rho+p\ge 0$) is satisfied
throughout the wormhole spacetime. In fact, for the solutions in 1) 2)
and 3), all energy conditions (the weak energy condition, the null energy condition,
the dominant energy condition, the strong energy condition, etc \cite{vis95,haw73})
are satisfied all the time and everywhere.

All the solutions presented above are dynamic. If we let $\dot{a} = \ddot{a}
=0$ in Eq.~(\ref{eq1}) and Eq.~(\ref{eq2}), we get $\rho + p = -(4\pi a^2)^{-1}
<0$, so there are no static solutions satisfying the weak energy condition.

Since the weak energy condition is satisfied and the spatial curvature is negative,
the wormhole spacetime constructed above expands forever if it expands initially.
Such a wormhole spacetime has no event horizons and no singularities in the future,
so an observer can travel from one side to the other if he travels to the
future. Though the spacetime may have singularities in the past as in
the big-bang case, there will be no problem if a time arrow exists as
in our universe. The time arrow will forbid any observer and light ray from
traveling to the past, then we only need to consider traveling to the
future. [The solutions in 1) and 4) even do not have singularities in the past, any
worldline in these solutions can be extended to the past infinitely.] In such
a wormhole spacetime with a time arrow,
an observer can travel from one side to the other. The observer will feel no
tidal force since the space is homogeneous. In fact he will not feel any
difference in local spatial geometry from the place where
he starts. If $\rho\ge0$ and $\Lambda\ge0$ --- as in the cases for our solutions,
the observer even will not feel the existence of the ``throat", since the wormhole
expands so rapidly that as the observer passes through the wormhole to
the other side he always sees that the cross-section ${\cal S}$ gets larger and
larger. To see this, let's consider a bunch of light rays passing through the
wormhole. Suppose the light rays
travel in such a way that at any moment their velocity is perpendicular to
${\cal S}$, i.e., along every light ray we have $\xi = {\rm constant}$ and
$\phi = {\rm constant}$. (This can be realized due to the symmetry of the
spacetime.) Let's take the tangent vector of the null geodesics of the light
rays to be
\begin{eqnarray}
    k^a = {1\over a}\left[\left({\partial\over\partial t}\right)^a + {1\over a}
          \left({\partial\over\partial\chi}\right)^a\right].
    \label{tang}
\end{eqnarray}
It can be checked that $k^a$ satisfies the affine geodetic equation $k^b\nabla_b
k^a = 0$, so $\lambda$ is the affine parameter of the null geodesics if $k^a =
(\partial/\partial\lambda)^a$. Further more, it can be shown that
\begin{eqnarray}
    \lambda = \int a dt\,.
    \label{aff}
\end{eqnarray}
The light rays form a null geodetic congruence
orthogonal
to ${\cal S}$. The expansion of the light rays, $\nabla_a k^a$, is related to
the variation of ${\cal A}$ (the area of ${\cal S}$) along the light rays by
$d\ln{\cal A}/d\lambda = \nabla_a k^a$ \cite{mtw73}. So we have
\begin{eqnarray}
    {d\ln{\cal A}\over d\lambda} = {2\over a^2}\left(\tanh\chi + \dot{a}\right).
    \label{varea}
\end{eqnarray}
Eq. (\ref{varea}) can also be obtained from Eq. (\ref{area}) and Eq. (\ref{aff})
by using the fact that $d\chi = dt/a$ along the light rays.
From Eq.~(\ref{eq1}), if $\rho$ and $\Lambda$ are non-negative, $\dot{a}^2\ge 1$
always. If $\dot{a}>0$ --- which is true for the expanding solutions described
above, we always have $\dot{a}\ge 1$. Since $\tanh\chi>-1$, we have
$\tanh\chi + \dot{a} > 0$ all the time. So, along the light rays,
$d\ln{\cal A}/d\lambda$ is always positive. This must also be true
for the observer since he can never travel faster than light: as the observer
travels through the wormhole, he will see that the size of ${\cal S}$ keeps
increasing.

Hochberg and Visser have demonstrated that violation of the
null energy condition is a generic feature of all wormholes, no matter they
are static or dynamic \cite{hoc98,hoc98a}. (The null energy condition requires only
$\rho+p\ge 0$, so the weak energy condition implies the null energy condition.
In other words, if the null
energy condition is violated the weak energy condition must be violated too.)
Their arguments essentially rely on their definition of wormhole throats and their 
assumption that all wormholes have throats with their definition.
They define a wormhole throat to be a marginally anti-trapped surface, which means
that bundles of light rays that enter the wormhole at one mouth and emerge from
the other must have cross-sectional areas that initially decrease and then increase
\cite{hoc98,mor88a}. With this definition of throats, the wormhole spacetime
constructed in this paper does not have a
throat if $\rho$ and $\Lambda$ are non-negative, since bundles
of light rays propagating through it always diverge.
However, the spacetime I have constructed exhibits a well-defined wormhole
structure with a well-defined throat (with the ordinary meaning \cite{mor88,mor88a})
on the spatial hypersurface at any moment: two large spaces are connected by a
narrow tube. My results do not conflict with the claim of
Hochberg and Visser because their definition of wormhole throats does not apply
to my case. This means that Hochberg and Visser's definition of wormhole
throats cannot cover all kinds of wormhole spacetimes, there exist spacetimes
which have unambiguous wormhole structures but do not have throats with
Hochberg and Visser's definition. (A
classical example of wormhole that cannot be covered by Hochberg and Visser's
definition is the Einstein-Rosen bridge \cite{ein35}. As bundles of light rays
pass through the Einstein-Rosen bridge from one side, their cross-sectional areas
keep decreasing until the light rays  hit the future spacetime singularity.)

In summary, I have constructed a wormhole spacetime which has many nice features.
The spacetime represents two {\it open} universes connected by a wormhole.
Since more and more observations show that we are living in a universe with
a low mass density, it is quite possible that our universe
is open (\cite{got97,got98} and the references therein; certainly there is another
possibility that we are living in a flat universe with a positive cosmological
constant). Maybe one of the open universes in the wormhole spacetime I have 
constructed is just the universe that we are living in, at least there is no 
observational
evidence to exclude this possibility! It would be interesting to investigate
the fluctuations of the cosmic microwave background in such a wormhole spacetime
to see if it is possible to tell if we are really living in such an open universe
\cite{cor98,cor00}. The wormhole spacetime can exactly solve the Einstein equations with
a stress-energy tensor satisfying the weak energy condition. The solutions
presented in the paper are {\it true} solutions to the Einstein equations since we
are aware of the matter contents required by these solutions, which are dust,
radiation, vacuum, or a positive cosmological constant. In contrast, for most
wormhole solutions in literatures (the Einstein-Rosen
bridge is a prominent exception), a geometry is assumed and the
stress-energy tensor is calculated from the metric by solving the Einstein
equations reversely.

I am grateful to J. Richard Gott for helpful discussions.
This work was supported by the NSF grant AST-9529120.

\end{document}